\documentclass{article}
\usepackage{spconf,amsmath,graphicx, booktabs, multirow}

\title{Impact of visual assistance for automated audio captioning}
\name{Wim Boes, Hugo Van hamme\thanks{This work was supported by Research Foundation Flanders (FWO) under grant S004923N of the SBO programme and the Flemish Government under ``Onderzoeksprogramma AI Vlaanderen''.}}
\address{
ESAT, KU Leuven, Belgium}
\begin{document}
\ninept
\maketitle
\begin{abstract}
We study the impact of visual assistance for automated audio captioning. Utilizing multi-encoder transformer architectures, which have previously been employed to introduce vision-related information in the context of sound event detection, we analyze the usefulness of incorporating a variety of pretrained features.

We perform experiments on a YouTube-based audiovisual data set and investigate the effect of applying the considered transfer learning technique in terms of a variety of captioning metrics.

We find that only one of the considered kinds of pretrained features provides consistent improvements, while the others do not provide any noteworthy gains at all. Interestingly, the outcomes of prior research efforts indicate that the exact opposite is true in the case of sound event detection, leading us to conclude that the optimal choice of visual embeddings is strongly dependent on the task at hand.

More specifically, visual features focusing on semantics appear appropriate in the context of automated audio captioning, while for sound event detection, time information seems to be more important. 
\end{abstract}
\section{Introduction}
\label{sec:intro}

Sound recognition has recently become a popular machine learning discipline. This is largely thanks to the organization of yearly editions of the Detection and Classification of Acoustic Scenes and Events (DCASE) challenge~\cite{DCASE2016proceedings, DCASE2017proceedings, DCASE2018proceedings, DCASE2019proceedings, DCASE2020proceedings, DCASE2021proceedings}. This has in turn been facilitated by the publishing of relevant audio data sets, such as Audio Set~\cite{gemmeke2017audio}, which is useful in the context of many related subtasks.

Sound recognition encompasses many partially connected problems, such as classification and/or spatial localization of acoustic events. In this work we specifically deal with the relatively complex task of automated audio captioning, of which the objective is to generate informative transcriptions of auditory recordings.

The last three versions of the DCASE challenge, task 6 (or at least a subproblem thereof) has been devoted to automated audio captioning~\cite{DCASE2020proceedings, DCASE2021proceedings}. In all of these editions, Clotho~\cite{drossos2020clotho} was used for training and evaluation of the submitted systems. This data set consists of samples extracted from Freesound, a large-scale sound library, accompanied with crowdsourced captions.

Models employed for tackling this task and data typically consist of a mixture of convolutional, recurrent, and attention-based (e.g., transformer~\cite{vaswani2017attention}) neural networks organized in an encoder-decoder structure. The best-performing systems usually also utilize some additional mechanisms that provide performance improvements, such as sentence length estimation~\cite{koizumi2020ntt}, keyword-based extensions~\cite{koizumi2020ntt, yuan2021dcase}, pretraining~\cite{yuan2021dcase, xu2022sjtu} and reinforcement learning~\cite{xu2022sjtu}.

Clotho~\cite{drossos2020clotho} is a commonly utilized collection for automated audio captioning, but it is not the only one. An alternative designed for a similar purpose is AudioCaps~\cite{kim2019audiocaps}, which consists of a subset of the recordings in AudioSet~\cite{gemmeke2017audio}, accommodated with crowdsourced captions. In contrast to the previously described data option, the associated clips originate from YouTube, which opens up possibilities in terms of multimodal processing, as visual information can obviously be readily extracted in this case.

The merit of incorporating the visual modality has recently started to be investigated for text-based audio retrieval (and its inverse problem), for which the objective is to select sound samples from a pool of candidates that best fit a written query. In~\cite{koepke2022audio}, it is shown that in this context, including visual features can certainly lead to improvements. This endeavor is clearly related to task at hand, but is considerably less complex, as it does not involve natural language generation. As a consequence, in this work, we attempt to extend the current state of research by also analyzing the usefulness of vision-related information for automated audio captioning.

To this end, we start from the multi-encoder transformer architectures described in \cite{boes2022multi}. They have been utilized successfully for incorporating visual features in the context of audio tagging and sound event detection, which involve classification of auditory clips. We apply adaptations in order for them to become suitable to the problem at hand, i.e., automated audio captioning. This is relatively simple, as the base system is already relatively similar to many models employed in the context of the target task, such as some of the winners of the DCASE challenge previously mentioned~\cite{yuan2021dcase, xu2022sjtu}.

In Section~\ref{sec:meth}, we dive further into the employed methodology, describe the employed models and elaborate upon how they are used to assess the usefulness of visual information in the context of automated audio captioning. Next, in Section~\ref{sec:setup}, we provide more specific details on the experimental setup, and in Section~\ref{sec:res}, we report the results of all performed trials and analyze them. Finally, in Section~\ref{sec:conclusion}, we summarize the most important conclusions.


\section{Method}
\label{sec:meth}

In this section, we elaborate upon the used method. More specifically, we explain the models employed for assessing the usefulness of visual information in the context of automated audio captioning. They are modified versions of the multi-encoder transformer systems for visually assisted sound event detection proposed in \cite{boes2022multi}.

\subsection{Architecture}

\begin{figure}[!ht]
\centering
\includegraphics[scale=0.75]{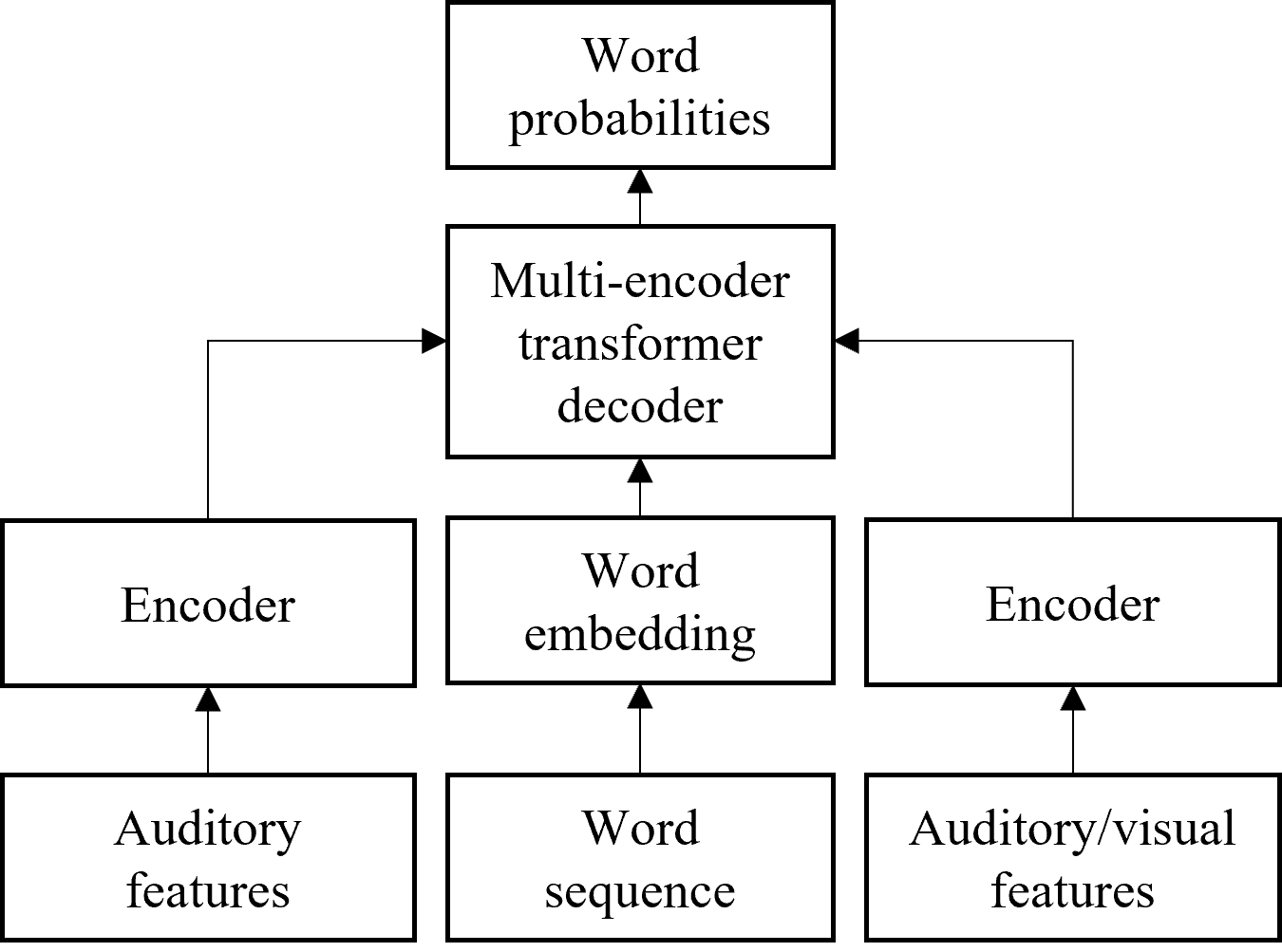}
\caption{Schematic representation of the employed model}
\label{fig:model}
\end{figure}

A schematic visualization of the architecture is given in Figure~\ref{fig:model}.

Two series of features are processed by an equal amount of encoders. The first sequence always consists of auditory inputs, while the second can either be acoustic or visual in nature. More information on the specific types of embeddings employed in this work are given in Section~\ref{sec:setup}. The main goal of this project is to investigate the merit of vision-related information in the context of automated audio captioning, and not necessarily to achieve hyperoptimized performance. Hence, the encoder structure is kept relatively simple: It consists of a single feedforward layer with ReLU activation mapping the embeddings to 128-dimensional vectors. This choice is further justified by the fact that all of the considered features are already quite informative as they are all extracted from pretrained models. 

The component combining the outputs of the two encoders and utilizing them to generate captions of the provided clips is the multi-encoder transformer decoder~\cite{boes2022multi, lohrenz2021multiencoder}. It is almost equal to the originally proposed variant in~\cite{vaswani2017attention} (or more accurately, the slightly modified version described in~\cite{xiong2020layer}), but there is one difference: All multi-head cross-attention blocks are duplicated and used to produce two sets of intermediate features, one for each series of vectors produced by the two encoders. These embeddings are linearly mixed and the result can be processed regularly further into the system. Naturally, the choice of the interpolation weights becomes an important hyperparameter consideration as a result and is discussed in Section~\ref{sec:setup}. 

The transformer decoder consists of 2 layers, each of them using attention modules with 4 heads producing 128-dimensional outputs. The feedforward layer with ReLU activations in these components project their inputs to 2048 vectors. Dropout~\cite{srivastava2014dropout} is employed at the appropriate places (as in the original work~\cite{vaswani2017attention}) with a rate of 0.2.

Before being fed to the decoder module, the textual inputs provided during learning as well as evaluation are first converted into real-valued vectors by means of a word embedding layer. This component is pretrained (but not fixed) using a continuous bag-of-words model~\cite{mikolov2013efficient} optimized on the training partition of the considered data set described in Section~\ref{sec:setup}. Furthermore, sinusoidal positional encodings are added to the token features to incorporate ordering information into the system, as also proposed in~\cite{vaswani2017attention} and \cite{gehring2017convolutional}.

\subsection{Postprocessing}

During training, the reference caption(s) are available and no further postprocessing is necessary to perform learning. On the other hand, in the evaluation phase, this is obviously not the case, and extra steps are necessary to properly produce transcriptions. More specifically, a decoding procedure has to be incorporated: The words are iteratively generated based on the output probabilities of the considered model. Token selection is done using beam search with a width of 5, a max depth of 20 and a length normalization factor of 1.

\section{Experimental setup}
\label{sec:setup}

In this section, the setup used during all of the performed experiments is discussed. More specifically, we go into the employed data and corresponding preprocessing methods, detail hyperparameters used in the training phase and elaborate on the captioning metrics utilized for evaluation of the considered models.

\subsection{Data}

In this project, we utilize the AudioCaps data collection~\cite{kim2019audiocaps}. It is a subset of the well-known AudioSet~\cite{gemmeke2017audio}, which features over two million YouTube clips of (mostly) 10 seconds long and is accompanied with an ontology of 632 environmental sound event classes. The considered selection is also provided with textual captions and can therefore be used to tackle automated audio captioning.

Because the clips in the collection originate from YouTube, we are able to extract visual features on top of the usual auditory embeddings and accomplish our objective, i.e., investigate the usefulness of vision for the task at hand. Unfortunately, this data set is not carefully curated, and the only way to obtain the samples is to perform manual download, which inevitably leads to some availability issues. 

The resulting data set used in this work is split into three disjoint partitions for training, validation and evaluation. They contain 45964, 459 and 907 clips respectively. Each sample in the latter two subcollections are provided with five captions, while the examples in the subset for learning only come with a single textual description.

\subsection{Preprocessing}

In this work, we use a number of different pretrained auditory and visual features for generating captions. In this section, we detail all of the preprocessing steps performed to obtain these embeddings. Interesting to note is that the choice of vision-related options is the same as in the work on multi-encoder attention-based architectures for visually assisted audio tagging and sound event detection~\cite{boes2022multi}, allowing us to broaden the analysis in Section~\ref{sec:res} by making interesting comparisons between different sound recognition tasks. 

\subsubsection{PANN auditory features}

To obtain pretrained auditory features, we first resample the audio clips to 44.1 kHz. Then, log mel magnitude spectrograms with 64 frequency bins are extracted using a Hamming window of 1024 samples (corresponding to about 25 ms) and a hop length of 431 samples (corresponding to about 10 ms). For a recording of 10 seconds, this results in 1024 frames. Next, these spectral maps are fed to a version of the CNN10 PANN~\cite{kong2020panns}, pretrained for audio tagging on the large-scale AudioSet data set~\cite{gemmeke2017audio}. The 512-dimensional outputs of the last feedforward layer of this neural network are used as embedding sequences in this project. The combination of all these steps leads to a series of 64 vectors for an acoustic input that is 10 seconds long.

These auditory features have previously been employed to great success in the context of automated audio captioning, such as in~\cite{mei2021encoder}.

In contrast to the other pretrained models mentioned in this section, the CNN10 PANN network~\cite{kong2020panns} is included into the complete architecture and its parameters are allowed to update during learning. 

\subsubsection{OpenL3 visual features}

OpenL3~\cite{cramer2019look} is an embedding model trained in a self-supervised
manner to predict correspondence between auditory and visual streams. It is pretrained on  the large-scale Audio Set~\cite{gemmeke2017audio}.

To obtain pretrained visual features, still frames are first sampled from the visual clips at a rate
of about 6.5 fps. The images are fed into the video subnetwork of OpenL3~\cite{cramer2019look}. For a recording of 10 seconds, these steps lead to a series of 64 512-dimensional vectors.

\subsubsection{Temporally coherent (TC) visual features}

To obtain pretrained visual features, still frames are first sampled from the visual clips at a rate
of about 6.5 fps. The images are fed into the video embedding model described in~\cite{knights2021temporally}, pretrained on various action recognition data sets in a self-supervised way using a loss enforcing temporal coherency. For a recording of 10 seconds, these steps lead to a series of 64 2048-dimensional vectors.

\subsubsection{VGG16 visual features}

To obtain pretrained visual features, still frames are first sampled from the visual clips at a rate
of about 6.5 fps. The images are fed into VGG16~\cite{simonyan2015very}, a convolutional network for image classification, pretrained on the ImageNet data set~\cite{russakovsky2015imagenet}. For a recording of 10 seconds, these steps lead to a series of 64 4096-dimensional vectors.

\subsection{Data augmentation}

As mentioned before, the goal of this project is to analyze the merit of including visual information in the context of automated audio captioning, and not to obtain ideal performance. Therefore, we only utilize a limited amount of data augmentation to prevent overfitting of models. More specifically, we solely employ SpecAugment~\cite{park2019specaugment}: Time and frequency masking algorithms are applied with maximum widths of 64 and 8 respectively to the log mel spectrograms fed as inputs to the pretrained PANN networks~\cite{kong2020panns} described earlier.

\subsection{Training and evaluation}

All of the models experimented with in this work are trained and evaluated using the PyTorch~\cite{paszke2019pytorch} and Gensim~\cite{rehurek2010software} toolkits.

\subsubsection{Training}

As mentioned in Section~\ref{sec:meth}, special attention has to be given to the interpolation hyperparameters used to combine the intermediate features computed based on the outputs of the two encoders, associated with (potentially) different inputs. During the learning phase, they are chosen randomly in the way suggested in~\cite{boes2022multi}: The mixing weight of the first encoder, which always takes in acoustic features, is sampled from a uniform distribution between 0.25 and 1 per batch. As audio is generally speaking the most salient modality for the task at hand, this value is not allowed to go all the way down to 0. Naturally, the weight of the second encoder, which might take in visual embeddings, is set in such a way that the sum of both equals 1.

All models are trained for 30 epochs, and each batch presented to the systems during the learning phase contains 8 samples. 

Adam~\cite{kingma2014adam} is utilized for optimization of the model parameters, based on the categorical cross entropy loss function. Learning rates are ramped up linearly from 0 to 0.001 for the first 5 epochs. Subsequently, they decay with a factor of 0.1 after every 7 epochs.

\subsubsection{Evaluation}

For evaluation, interpolation weights are optimized on validation data. The range between 0 and 1 is explored in increments of 0.05. The final decision is based on the METEOR score, which is explained in what follows. Empirically, we find that this choice also leads to near-optimal performance for the other metrics listed below.

For the evaluation of automated audio captioning, several metrics can be utilized. In most cases, they are borrowed from natural language processing tasks such as machine translation. We refer the reader to~\cite{kilickaya2017re} for an overview of the measures which are most commonly used in this context. In what follows, we provide elementary explanations of the scores which are employed in this work.

BLEU-$4$~\cite{papineni2002bleu} is based on modified precision scores of $n$-grams in the captions produced by the analyzed model, up to order 4.

METEOR~\cite{banerjee2005meteor} is computed as a weighted harmonic mean of the precision and recall scores of word matches between captions, and includes stemming and synonym detection techniques.

ROUGE-L~\cite{lin2004automatic} is based on the length of the longest common subsequence between generated and reference captions.

CIDEr~\cite{vedantam2015cider} calculations are based on comparisons between $n$-grams in the generated and reference captions, and includes term frequency inverse document frequency weighting~\cite{robertson2004understanding}.

SPICE~\cite{anderson2016spice} computations are based on the similarity between scene-graph tuples of the generated and reference captions.

SPIDEr~\cite{liu2017improved} is calculated as the average of the aforementioned SPICE and CIDEr scores, and compromises between evaluation based on syntactical and semantic information.

All considered metrics are computed using the output probabilities of the models obtained after the last training epoch.

\section{Experimental results}
\label{sec:res}

In this section, we analyze the automated audio captioning scores listed in Section~\ref{sec:setup} obtained by the considered multi-encoder transformers using different types of auditory and visual features, as discussed in Section~\ref{sec:meth}. We report metrics which are averaged over 20 training runs with independent model parameter initialization instances to improve the reliability of the results. 

\begin{table*}[!ht]
\caption{Scores of multi-encoder transformers}
\label{tab:res}
\centering
{\setlength{\tabcolsep}{11.460pt}
\begin{tabular}{@{}llcccccc@{}}
\toprule
\textbf{Encoder} &
\textbf{Evaluation} &
\textbf{BLEU-}$\mathbf{4}$ &
\textbf{METEOR} & 
\textbf{ROUGE-L} &
\textbf{CIDEr} & 
\textbf{SPICE} &
\textbf{SPIDEr} \\
\textbf{inputs} &
\textbf{encoder weights} &
\textbf{(\%)} &
\textbf{(\%)} & 
\textbf{(\%)} &
\textbf{(\%)} & 
\textbf{(\%)} &
\textbf{(\%)} \\
\midrule
PANN auditory features & 0.5 & 24.45 & 22.52 & 46.35 & 65.32 &  16.41 & 40.87 \\
PANN auditory features & 0.5 & \textpm 0.20 & \textpm 0.060 & \textpm 0.071 & \textpm 0.35 & \textpm 0.054 & \textpm 0.12 \\
\midrule
PANN auditory features & 1 & 24.48 & 22.52 & 46.35 & 65.39 & 16.42 & 40.91 \\
OpenL3 visual features & 0 & \textpm 0.19 & \textpm 0.052 & \textpm 0.067 & \textpm 0.30 & \textpm 0.051 & \textpm 0.10 \\
\midrule
PANN auditory features & 0.75 & 24.48 & 22.55 & 46.40 & 65.57 & 16.48 & 41.02 \\
TC visual features & 0.25 & \textpm 0.18 & \textpm 0.056 & \textpm 0.081 & \textpm 0.39 & \textpm 0.057 & \textpm 0.14 \\
\midrule
PANN auditory features & 0.75 & \textbf{24.78} & \textbf{22.85} & \textbf{46.85} & \textbf{66.87} & \textbf{16.91} & \textbf{41.89} \\
VGG16 visual features & 0.25 & \textpm 0.15 & \textpm 0.050 & \textpm 0.076 & \textpm 0.33 & \textpm 0.055 & \textpm 0.11 \\
\bottomrule
\end{tabular}}
\end{table*}

Table~\ref{tab:res} contains the scores and associated standard deviations of the models incorporating different pretrained features. As mentioned in Section~\ref{sec:meth}, the first encoder always takes in auditory embeddings, while the other inputs are variable. The interpolation weights used at inference time are also listed. The results are competitive with the current state of the art as catalogued in a recent survey~\cite{mei2022automated}.

The multi-encoder transformers with OpenL3 and TC visual features clearly do not significantly
outperform the variant only employing auditory inputs. On the other hand, the systems incorporating VGG16-based embeddings do achieve better performance. In what follows, we will be diving deeper into this interesting outcome.

Firstly, it is necessary to discuss the encoder weights employed during evaluation. For the model only using auditory inputs, these do not make any real difference for obvious reasons. For the variant with OpenL3 features, the interpolation hyperparameter associated with the visual stream is set to 0, which means that the optimal choice is to ignore the knowledge encapsulated by these vectors. For the system employing the other considered pretrained vision-based embeddings, this is not the case, however, the contribution of the acoustic information still dominates as the connected mixing hyperparameter remains relatively large. This makes sense, as audio is clearly the most salient modality for the task at hand.

\begin{figure}[!ht]
  \centering
  \medskip
  \includegraphics[scale=0.75]{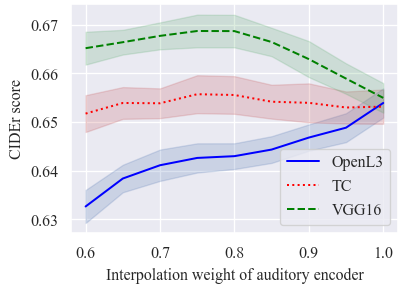}
  \caption{CIDEr scores of audiovisual multi-encoder transformers}
  \label{fig:plot}
\end{figure}

When it comes to the visual features, as previously noted, only the VGG16 embeddings significantly improve the performance of the models. A further visualization of the results is given in Figure~\ref{fig:plot} for CIDEr scores, clearly demonstrating the disparity in response when the interpolation weight of the auditory encoder is lowered.

Table~\ref{tab:resvis} lists a subset of results obtained by the audiovisual multi-encoder transformers when the interpolation weight of the acoustic stream is forced to 0 during inference, and only visual information can be utilized to generate captions. As expected, all models strongly underperform when comparing against the outcomes in Table~\ref{tab:res}, but again, we observe that VGG16 features perform the best.

\begin{table}[!ht]
\caption{Vision-only scores of multi-encoder transformers}
\label{tab:resvis}
\centering
{\setlength{\tabcolsep}{4.675pt}
\begin{tabular}{@{}lccccc@{}}
\toprule
\textbf{Visual} &
\textbf{BLEU-}$\mathbf{4}$ &
\textbf{METEOR} & 
\textbf{ROUGE-L} &
\textbf{CIDEr} & 
\textbf{SPICE} \\
\textbf{features} &
\textbf{(\%)} &
\textbf{(\%)} & 
\textbf{(\%)} &
\textbf{(\%)} & 
\textbf{(\%)} \\
\midrule
OpenL3 & 6.630 & 11.41 & 26.81 & 8.523 & 4.952 \\
\midrule
TC & 8.401 & 12.17 & 29.35 & 15.57 & 6.050 \\
\midrule
VGG16 & \textbf{10.23} & \textbf{13.69} & \textbf{32.26} & \textbf{22.22} & \textbf{7.976} \\
\bottomrule
\end{tabular}}
\end{table}

In~\cite{boes2022multi}, it is shown that incorporating OpenL3 and TC features into multi-encoder transformers is beneficial for sound event detection, while VGG16 embeddings provide little merit. Curiously, this is the opposite of what we find in this work on audio captioning. 

This can be explained by differences with regard to the visual information captured by the pretrained vectors. For sound event detection, which requires segmentation, time-related details are important, and thus, features extracted from models built for tasks with temporal facets are suitable. In contrast, in the case of automated audio captioning, semantic details are essential and it is more appropriate to use embeddings focusing on this knowledge, such as those produced by VGG16, which was designed for object detection.

\section{Conclusion}
\label{sec:conclusion}

In this work, we investigated the impact of incorporating pretrained visual features into models for automated audio captioning. To this end, we employed multi-encoder transformer systems, which have previously been utilized to add vision-related information in the context of classification-based sound recognition tasks to great effect.

Experiments were performed using a YouTube-based audiovisual data set and a variety of pretrained visual features. Performance was measured in terms of multiple captioning metrics.

We showed that the inclusion of just one of the three examined sets of pretrained visual features led to significant performance improvements for automated audio captioning. Fascinatingly, this outcome is the direct opposite of the conclusions formulated in prior related research on audio tagging and sound event detection: For those classification-based tasks, only the incorporation of the other regarded types of vision-related embeddings proved to be useful.

This phenomenon can be explained by the difference in knowledge encapsulated by the inspected pretrained visual inputs. Features focusing on time-related aspects seem to be useful for sound event detection, which makes sense as this task involves a temporal segmentation aspect. However, in the context of automated audio captioning, these details are largely irrelevant and it is more appropriate to employ embeddings concentrating on semantic information, such as vectors extracted from models designed for object detection.

For future research, it could be wise to investigate the impact of incorporating pretrained visual features into models for other sound recognition tasks. It might also be useful to design embeddings combining the strengths of those examined in this work, i.e., create vectors encapsulating relevant temporal as well as semantic information. 

\bibliographystyle{IEEEbib}
\bibliography{refs_shortened}

\end{document}